\documentclass[a4paper]{jpconf}
\usepackage{graphicx}

\usepackage{citesort,amsmath,amssymb}

\def\la{\langle}
\def\ra{\rangle}

\def\be{\begin{equation}}
	\def\ee{\end{equation}}
\def\la{\langle}
\def\ra{\rangle}

\def\Cas{\mathrm{Cas}}

\newcommand{\beq}{\begin{eqnarray}}
	\newcommand{\eeq}{\end{eqnarray}}
\def\be{\begin{equation}}
	\def\ee{\end{equation}}
\newcommand\eq[1]{Eq.~(\ref{#1})}

\newcommand\reff[1]{(\ref{#1})}

\newcommand{\EE}{\mathcal E}

\begin{document}

%\title{Fluctuation-induced Interactions in Micro- and Nano-systems: Survey of Analytical Results for Basic Models}

\title{Fluctuation-induced Interactions in Micro- and Nano-systems: Survey of Some  Basic Results}

\author[DMD]{Daniel Dantchev$^{1,2}$}

\address{$^1$ Institute of
	Mechanics, Bulgarian Academy of Sciences, Academic Georgy Bonchev St. building 4,
	1113 Sofia, Bulgaria}
\address{$^2$ Max-Planck-Institut f\"{u}r Intelligente Systeme, Heisenbergstrasse 3, D-70569 Stuttgart, Germany} 
%\address{
%		$^3$ IV. Institut f\"{u}r Theoretische Physik, Universit\"{a}t Stuttgart, Pfaffenwaldring 57, D-70569 Stuttgart, Germany}

\ead{danieldantchev@gmail.com}

\begin{abstract}
On the examples of the quantum-electrodynamical Casimir force, as well as critical Casimir and Helmholtz forces,  we present a review of some  results available for the class of fluctuation induced forces.  In addition, we also concisely  present examples of other such fluctuation-induced forces.  On the instance of the Ising model we discuss the connection between the Casimir and Helmholtz forces. We discuss the importance of the presented results for the nanotechnology, and especially for devising micro- or nano-systems, and for their assembly. Some important problems for the nanotechnology, following from the currently available experimental findings, are spelled out and possible strategies for their overcoming are outlined.  
\end{abstract}

\section{Introduction}

The confinement of a fluctuating field by the surfaces of some material bodies induces forces acting on the confining surfaces of these bodies.
These forces go under the general name of fluctuation induced forces.

Let us note that fluctuations are ubiquitous: they unavoidably appear in any matter either due to its quantum nature or due to nonzero
temperature of the material bodies and of the confined medium. In addition, any of these bodies and the medium can be at different temperatures thus creating a set of non-equilibrium phenomena. The bodies can also be in motion with
respect to the medium or each other. All of the above options generate a plethora of possible fluctuation induced forces. 

Naturally, the  strength of the fluctuation-induced forces is  proportional to the driving energy of the fluctuations, and
thus to Planck’s constant $h$ in quantum systems and to temperature $T$ in \textit{i)} classical systems, and in quantum systems \textit{ii)} when $k_B T \gg  h \nu_0$, where $\nu_0$ is some characteristic frequency of the quantum system. 

Immersing bodies of given shapes and materials into a fluid medium changes its fluctuation spectrum. This change has to be
in accord with the geometry, the relative positions and orientations, and the materials properties of the bodies. Thus, if these
fluctuations are correlated in space, the dependence of their spectrum on the relative positions and orientations of the
bodies generates an effective force and torque, respectively, acting between them. If the excitations of the fluctuations lack
an energy gap, as it is the case, e.g., for photons, Goldstone bosons, and the fluctuations of an order parameter at criticality,
the fluctuation induced force acquires an algebraic decay and, thus, becomes \textit{long-ranged}.

When the degrees of freedom can enter and leave the region between the interacting
objects one speaks about \textit{Casimir force}. In the case of the electromagnetic Casimir force the medium is the vacuum, and the underlying mechanism is the set of quantum zero point or temperature fluctuations of the electromagnetic field. Then one speaks about the \textit{quantum mechanical} (QED) Casimir effect.
Investigations devoted to that are currently performed on many fronts of research ranging from attempts to unify
the four fundamental forces of nature \cite{MT97,M2001,M2004} to rather
practical issues such as the design and the performance of micro- and nano-machines \cite{GLR2008,KMM2011,RCJ2011,FAKA2014,FMRA2014}.

 The now widely-investigated \textit{critical Casimir force} (CCF) results from the fluctuations of an order parameter and more generally the thermodynamics of the medium supporting that order parameter in the vicinity of a critical point. Recently, a review on the exact results available for the CCF has been published in Ref. \cite{DD2022}.

In recent Letter  \cite{DR2022} we have introduced the terms of a \textit{Helmholtz fluctuation induced force} (see also \cite{rem}). It is a force in which the order parameter value is fixed. We stress, that in customarily considered applications of, say, the equilibrium Ising model to binary alloys or binary liquids, if one insists on full
rigor, the case with order parameter fixed must be addressed. In \cite{DR2022}  via deriving there exact results on the example of Ising chain with fixed magnetization and under periodic boundary conditions, we have shown that the Helmholtz force has a behavior very different from that of the Casimir force in the same model and under the same boundary conditions. It is interesting to note that, actually, the studied Helmholtz force has a behavior similar to the one appearing in some
versions of the big bang theory --- strong repulsion at high
temperatures, transitioning to moderate attraction for intermediate
values of the temperature, and then back to repulsion,
albeit much weaker than during the initial period of highest
temperature.  

We stress that the definition and existence of Helmholtz force is by no means limited to the Ising chain and can be addressed, in principle, in any model of interest - see, e.g., Refs. \cite{RSVG2019,GVGD2016,GGD2017}.

We note that the issue of the ensemble dependence of fluctuation-induced forces pertinent to the ensemble has yet to be fully studied in different models. Indeed, it can be studied in  \textit{any} ensemble and in \textit{any} model defined within that ensemble. 

In the current text,  we review some recent and present some new both exact and numerical results for the behavior of the Casimir and Helmholtz force.  We find that all significant results are consistent with the expectations of finite size scaling theory. 

The structure of the review is as follows. First, in Sect. \ref{sec:examples} we present examples of a variety of fluctuation-induced effective forces. This is aimed to indicate how broad and diverse is the scientific field associated with such forces. We will make an emphasis, however, on the quantum, see Sect.  \ref{sec:QED-Casimir}, and critical Casimir effects, see Sect. \ref{sec:CCE}, on which most of the theoretical efforts are concentrated. Then Sect. \ref{sec:Casimir-versus-Helmholtz} recalls the definitions of the Casimir and the Helmholtz forces in grand canonical, and canonical ensembles, respectively. Sect. \ref{sec:Helmholzt-force}, following Ref. \cite{DR2022}, presents some results available for the Helmholtz force, while in Sect. in  \ref{sec:canonical-versus-grand-canonical} we comment on the connection between the grand and the canonical ensembles in the Ising model case. The article ends with concluding comments and discussion in Sect. \ref{sec:conclusion}, with some emphasis on the problems existing in nanotechnology and on possible strategies for their overcoming.

\section{Examples of fluctuation-induced effective forces}
\label{sec:examples}

As already alluded  in the introduction, the confinement of a fluctuating field generates effective forces on the confining surfaces which are nowadays termed  fluctuation-induced forces.

\subsection{The example considered by Einstein}
We start by one example of such a type of a force  considered by Einstein. As early as in 1907   in his publication \cite{E07} he considered voltage fluctuations in capacitor systems which are due due to nonzero temperature $T$. Similar effects are also known to occur in  wires \cite{J28,N28}. For example, the famous Johnson-Nyquist formula tells that the mean-square noise current $\langle I^2\rangle$ depends on the resistivity $R$ and the temperature $T$ of a resistor according to  $\langle I^2\rangle=4k_B\,T\Delta f/R$, where $\Delta f$ is the measurement bandwidth.   Such fluctuations lead to forces which are of serious interest  for the operation of electro-mechanical devices \cite{RRJ2013} down-scaled to micro- or nano-levels.  
	
\subsection{The QED Casimir effect}	
\label{sec:QED-Casimir}

\begin{figure}[!htb]
		 	\begin{minipage}{0.495\columnwidth}
		 		\includegraphics[width=\columnwidth]{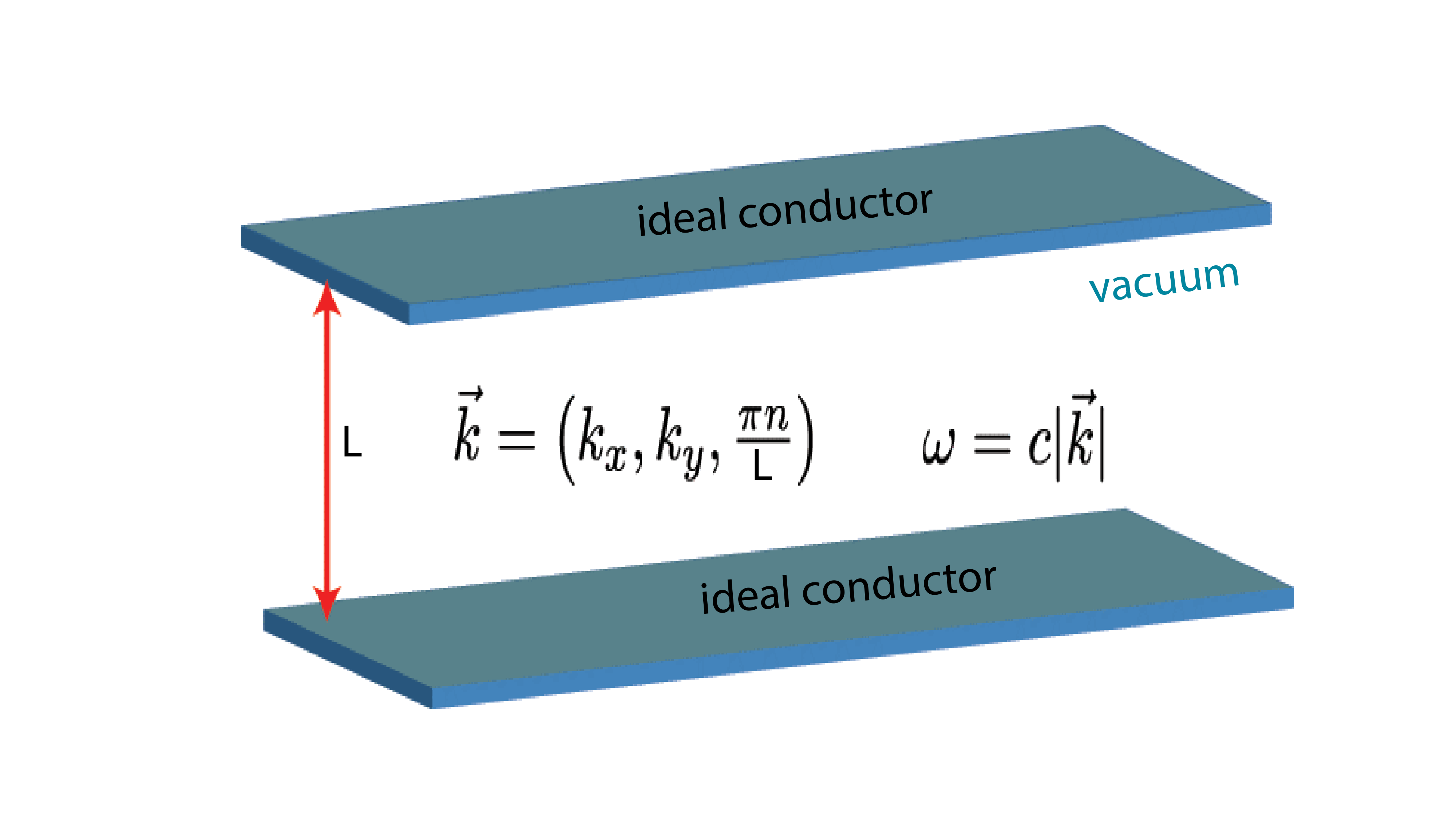}
		 		\caption{\label{Fig:QED} The set up of a system considered by Casimir in his original article \cite{C48}.}
		 	\end{minipage}\hspace{0.5pc}%
		 	\begin{minipage}{0.495\columnwidth}
		 		\includegraphics[width=\columnwidth]{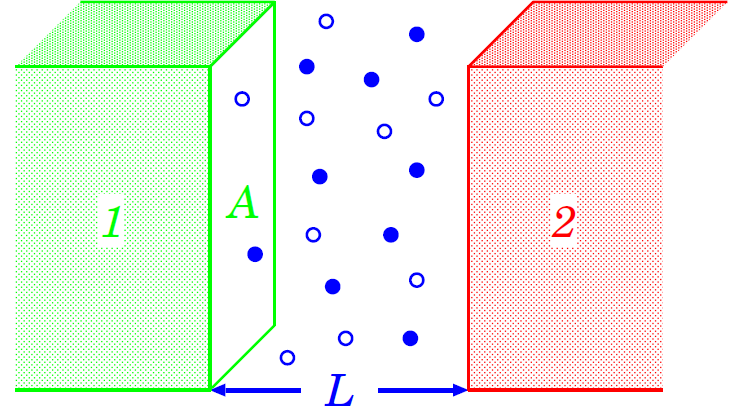}
		 		\caption{\label{Fig:TCE}The basic setup for discussing the thermodynamic Casimir effect \cite{FG78}.}
		 	\end{minipage} 
		 \end{figure}

The currently most prominent example of a fluctuation-induced force is the  one which we already alluded to and which is due to quantum or thermal fluctuations of the electromagnetic field.
		 These fluctuations  induce the so-called QED Casimir effect, which is named after the Dutch physicist H. B. Casimir, who first realized that in the case  of two perfectly conducting, uncharged parallel plates in vacuum and at $T=0$  (see Fig. \ref{Fig:QED}) these fluctuations lead to an attractive force between them \cite{C48}. 
		 Casimir demonstrated that the boundary conditions imposed by
		 two plates (in the
		 following denoted by $\|$) at a distance $L$ on the spectrum of the
		 quantum-mechanical zero-point fluctuations of the electromagnetic field lead to the remarkable mechanical effect of the appearance of a  long-ranged attractive force between the plates. For this force, divided by the cross-sectional area of the plates, he obtained
		 \begin{equation}\label{qCasimirf}
		 	F_{\rm Cas}^{\parallel}(L)=-\frac{\pi^2}{240}\frac{\hbar c}{L^4} = -1.3
		 	\times 10^{-3}\frac{1}{(L/{\rm \mu m})^4} {\rm \frac{N}{m^2}}.
		 \end{equation}
		 Intuitively the corresponding pressure can be viewed as the
		 difference in radiation pressure of virtual photons outside
		 and inside the pore formed by the two plates which results in an
		 attractive force $F_{\rm Cas}^{\parallel}$ between them \cite{MCG88}. 
		 
		 The relationship in \eq{qCasimirf} can be derived by considering the change of the structure of the electromagnetic modes between the two plates as compared with those in free space after assigning the zero-point energy $\frac 1 2 \hbar \omega$ to each electromagnetic mode, i.e., photon of frequency $\omega$. We emphasize that in the absence of charges on
		 the plates the mean value of both the electric  $\bf E$ and the magnetic field $\bf B$ vanishes, i.e.,
		 \begin{equation}
		 	\la {\bf E}\ra = 0 \qquad \mbox{and} \qquad \la {\bf B} \ra =0,
		 	\label{average}
		 \end{equation}
		 but
		 \begin{equation}
		 	\la {\bf E}^2 \ra \ne 0 \qquad
		 	\mbox{and} \qquad \la {\bf B}^2 \ra \ne 0,
		 	\label{dispersion}
		 \end{equation}
		 so that the expectation value of the energy due to the electromagnetic field, i.e., $\langle{\cal H}\rangle$ with 
		 \begin{equation}
		 	\label{EMFenergy}
		 	{\cal H}=\int  \left[\frac{1}{2}\varepsilon_0 \textbf{E}^2(\textbf{r})+\frac{1}{2\mu_0} \textbf{B}^2(\textbf{r})\right]\;d^3\textbf{r},
		 \end{equation}
		 is nonzero. In fact, according to
		 quantum field theory, the energy of the electromagnetic field in the vacuum state in free space is infinitely large; all physically relevant energies are measured relative to the energy of the vacuum. When the two, in the above sense  ideal, parallel metal plates are placed in free
		 space, the tangential component of the electric field and the normal component of the magnetic induction must vanish at the plate surfaces. As a result, not all zero-point oscillations occur. Subtracting from the energy of the allowed modes the ones of the vacuum
		 energy in free space, one obtains, after taking the derivative with respect to $L$, the result reported in \eq{qCasimirf}. In more details,  focusing on a single component of the electric field, which vanishes at the
		 surface of the metal (the discussion can be extended easily to
		 include all vector components), the simultaneous presence of two
		 parallel walls restricts the allowed wavevector $k_\bot$ of the field in the
		 direction normal to the plates, so that the field itself vanishes at both
		 walls (Dirichlet boundary conditions). No restriction, instead, is imposed on
		 the wavevector $k_\|$ parallel to the plates, which are assumed to have a large lateral extension with a transverse area $A$ and to be separated by a distance
		 $L$. Accordingly, one has $k_\bot = \pi n/L$, with $n=0,\pm 1, \pm 2,\cdots$, and the energy $\EE$ contained within the confined space takes the form
		 \be
		 \EE = \sum_{\bf k} \frac{1}{2}\hbar \omega(\bf k)
		 \label{eq:vacEn}
		 \ee
		 where the sum runs over all allowed values of ${\bf k}=({k_\|},k_\bot)$ and 
		 $\omega({\bf k})\equiv\omega_k=c|{\bf k}|= c \sqrt{k_\|^2 + k_\bot^2}$, with $|{\bf k}|=k$, is
		 the frequency of the mode with wavevector ${\bf k}$. As it stands,
		 the sum in Eq.~\reff{eq:vacEn} diverges due to the fact that $|\omega_k|$ grows as $k$ increases and because the sum lacks any ultraviolet cutoff, i.e., $k$ is allowed to grow unboundedly. In this context one is actually interested in the energy of the modes which can be ``contained'' by the metallic cavity; they are the only ones affected by the presence of the cavity itself. At sufficiently high frequencies the cavity becomes transparent to the
		 electromagnetic field. Note that within the assumptions made in deriving Eqs. \reff{qCasimirf} no materials-dependent properties enter.
		 Thus the force depends only on the Planck constant $\hbar$ and the speed of light in vacuum $c$, as well as on the geometry of the
		 pore (here characterized by $L$). The force does also not depend on the
		 electric charge $e$, implying that for the present conditions of the phenomenon the coupling of the electromagnetic field to matter is unimportant, such as any other interaction. Actually, under these premises, up to the numerical prefactor, \eq{qCasimirf} is fixed by \textit{dimensional analysis} alone. Indeed, the $L$-dependence of the force can easily be predicted on dimensional grounds: as any force $F_{\rm Cas}^\parallel$ it is $F_{\rm Cas}^\parallel \propto Energy/Length\propto E/L$. Then $E$ is proportional to the cause of the effect, i.e., the energy of the relevant fluctuations. At $T=0$, these  are the quantum fluctuations. Therefore, $E\sim h c/L$. If one normalizes per unite area $A\sim L^2$, the result is $F_{\rm Cas}^\parallel \propto h c/L^4$, which is the Casimir result cited in Eq. \eqref{qCasimirf}. 
		 
		 In the aftermath of Ref. \cite{C48}, there has been an intense theoretical effort to describe the force beyond the case of ideal plates by considering the actual dielectric properties of the two plates and of the medium in-between
		 \cite{L56,BG75,KMM2000,LR2000,BKM2000,GKM2002,EVM2003,TL2004,ES2004,P2006,CCJKMM2007,BKMM2009}.  Specifically one has to
		 mention the groundbreaking results of Lifshitz et al. \cite{L56,DLP61}, who developed a unified theory of the van der Waals and the Casimir  forces. Let us note that in discussion of the quantum Casimir effect the retarded van der Waals interactions are often called Casimir-Polder, or, shortly, Casimir interactions. However, one shall keep the notion of retarded van der Waals interactions  when discussing the thermodynamic Casimir effect in order to avoid confusion of these forces with the critical Casimir force which will be later at the center of our attention. 
		 
	Lifshitz et al. studied the case of two materials acting as walls and described by frequency-dependent dielectric permittivities $\varepsilon^{(n)}(\omega),n=1,2$, separated by a third material characterized by $\varepsilon^{(0)}(\omega)$. It turns out that in the limit of small separations (but still large compared with molecular scales) the Casimir force approaches the more familiar van der Waals force \cite{DLP61r,BKMM2009}. This more realistic description provides specific predictions which are amenable to high-precision measurements. The corresponding expressions for the force are also quite instructive in that it allows one to predict how materials properties of
		 the substances involved have to be tuned in order to modify the strength, and even the sign,  of the force.  According to Lifshitz, in order to calculate the Casimir
		 pressure for a set of three dielectric materials one needs to know their
		 permittivities along the imaginary frequency axis. Since experimental data for the complex permittivity
		 \begin{equation}
		 	\varepsilon(\omega)=\varepsilon^\prime(\omega)+ i
		 	\varepsilon^{\prime\prime}(\omega),
		 	\label{decomposition_of_perimittivities}
		 \end{equation}
		 where $\varepsilon^\prime, \varepsilon^{\prime\prime} \in {\mathbb R}$, exist only for  real frequencies, the permittivity along the imaginary frequency
		 axis has to be determined from the Kramers-Kronig relation (see, e.g.,
		 Ref. \cite{LPL84})
		 \begin{equation}
		 	\varepsilon(i\xi)=1+\frac{2}{\pi}\int_0^\infty \frac{x
		 		\varepsilon^{\prime\prime}(x)}{x^2+\xi^2}dx.
		 	\label{Kramers_Kronig}
		 \end{equation}
		 From Lifshitz theory one  can infer
		\cite{DLP61r} that there is
		the possibility to observe Casimir
		{\it repulsion} in the film geometry if the two  half-spaces (1) and (2) forming
		the plates and confining the film (0) exhibit permittivities which fulfill the
		relationship
		\begin{equation}
			\varepsilon^{(2)}(i\xi)<\varepsilon^{(0)}(i\xi)<\varepsilon^{(1)}(i\xi).
			\label{repulsion_condition}
		\end{equation}
		Experimentally repulsion occurs if the inequality in \eq{repulsion_condition}
		holds over a \textit{sufficiently wide
		frequency range}. Actually this is a widespread phenomenon shared by all
		substrate-fluid systems which show complete wetting, such as in the
		experiment by Sabisky and Anderson \cite{SA73}. Accordingly,
		Casimir repulsion is a common feature \cite{Di88} and has been already observed - see, e.g., Ref. \cite{MCP2009}.
		
		A standard approximation for obtaining the force between two bodies of nontrivial shape is the so-called Derjaguin approximation \cite{D34}.  It is know as Derjaguin approximation (DA) in colloidal science (see e.g. \cite{SHD2003} and p. 34 in \cite{BK2010}), and proximity force approximation in studies of QED Casimir effect (see e.g. p. 97 in \cite{BKMM2009}). The main idea behind the DA is that one can relate the knowledge for the interaction force/potential between two parallel plates with the one between two gently curved colloidal particles, when the separation between them is much smaller than the geometrical characteristics of the particles in question. More specifically, the DA states that in $d=3$ the interaction force $F^{R_{1},R_{2}}(L)$ between two spherical particles with radii $R_{1}$ and $R_{2}$ placed at a distance $L\ll R_{1},R_{2}$  is given by
		\begin{equation}\label{DASpSp}
			F_{\rm DA}^{R_{1},R_{2}}(L)=2\pi R_{\rm eff}\int_{L}^{\infty}f_{{\cal A}}^{\parallel}(z){\rm d}z,
		\end{equation}
		where $R_{\rm eff}^{-1}=R_{1}^{-1}+R_{2}^{-1}$ is an effective radius and $f_{{\cal A}}^{\parallel}$ -- force per unit area between parallel plates. When the sphere with radius $R_{1}\equiv R$ interacts with a plate one has $R_{2}=\infty$ and then \eref{DASpSp} is still valid with $R_{\rm eff}=R$.
		
		An improvement and generalization of the DA, called "surface integration approach" (SIA) has  been proposed in Ref. \cite{DV2012}. It has been used there to study van der Waals interactions between objects of arbitrary shape and a plate of arbitrary thickness. It delivers {\it exact }results if the interactions involved can be described by \textit{pair potentials}. The main advantage of this approach over the DA is that one is no longer bound by the restriction that the interacting objects must be much closer to each other than their characteristic sizes. The main result is that for the force acting between a $3\mathrm{d}$ object (say a colloid particle) $B\equiv\{(x,y,z),(x,y,z)\in B\}$ of general shape $S(x,y)=z$ and a flat surface bounded by the $(x,y)-$plane of a Cartesian coordinate system, one has
		\begin{eqnarray}\label{SIAgeneralsimple}
			F^{B,|}_{\rm SIA}(L) =\int_{A_{S}^{\rm to}}\int  f^{\parallel}_{{\cal A}}[{\rm S}(x,y)] \mathrm{d}x \mathrm{d}y
			-\int_{A_{S}^{\rm away}}\int  f^{\parallel}_{{\cal A}}[{\rm S}(x,y)] \mathrm{d}x \mathrm{d}y.
		\end{eqnarray}
		Eq. \eqref{SIAgeneralsimple} has a very simple intuitive meaning: in order to determine the force acting on the particle one has to subtract from the contributions stemming from surface regions $A_S^{\rm to}$ that "face towards" the projection plane  those from regions $A_S^{\rm away}$  that "face away" from it, where  $A_S^{\rm to}$ and $A_S^{\rm away}$  are the projections of the corresponding parts of the surface of the body on the $(x,y)-$plane. It is clear that if one takes into account only the contributions over $A_S^{\rm to}$  one obtains expression very similar to the DA. Both expressions in that case will differ only by the fact that while \eref{SIAgeneralsimple} takes into account that the force on a given point of the $S$ is {\it along the normal} to the surface at that point, the standard DA does not take this into account. Let us recall that \eref{SIAgeneralsimple} provides exact results for the interaction under the assumption that the constituents of the body interact via pair potentials. This is, strictly speaking, {\it not} the case of the force in QED Casimir, i.e., \eq{SIAgeneralsimple} is still an approximation. It is, however, clear that under mechanical equilibrium of the colloid in the fluid, the force is again along the normal to the surface at the point of the surface where it acts. Thus, one can get a reasonably good approximation to the effect of that force by keeping just the integration over part of the surface of the body that faces the plane. Let us note that the importance of the SIA approach has been already recognized and appreciated --- see, e.g., Refs.  \cite{RB2019,BR2019,Djafri2019,LF2022,EsquivelSirvent2023}. 
		 
	There is a vast amount of literature concerning the field of research of quantum Casimir effect. We just mention the review articles in Refs. 
	\cite{PMG86,MT88,LM93,MT97,M94,KG99,B99,BMM2001,M2001,M2004,L2005,KM2006,BW2007,GLR2008,BKMM2009,KMM2009,FPPRJLACCGKKKLLLLWWWMHLLAOCZ2010,CP2011,OGS2011,KMM2011,RCJ2011,MAPPBE2012,B2012,Bo2012,Ba2012,C2012,DGT2014,RHWJLC2015,KM2015c,SL2015,ZLP2015,WDTRRP2016,BEKK2017,WKD2021,GCMSM2021,Bimonte2022a}, recent studies of the dynamical Casimir effect (in which actual
	photons can be created if a single mechanical mirror undergoes accelerated
	motion in vacuum) \cite{M70,GK98,JJWF2009,FC2011,WJPJDND2011,NJBN2012,LPHH2013}, and studies of the effects which emerge in systems out of thermodynamic equilibrium (in which the material bodies are characterized by different temperatures) \cite{APSS2008,B2009,KEK2011,KEBK2011,MA2011,LBBAM2017,IF2021}. Currently the QED Casimir effect is a popular subject of research. The Casimir and Casimir-like effects are object of studies in
	quantum electrodynamics, quantum chromodynamics, cosmology, condensed matter physics, biology and, some elements of it, in nano-technology. The reader interested in that problematic can consult the existing reviews - see the above references to some of these reviews.

\subsection{The critical and the thermodynamic Casimir effects }
\label{sec:CCE}

The fluctuations of the order parameter describing a continuous phase
	transition of a many-body system leads, as explained above, to the so-called \textit{critical Casimir effect} \cite{FG78}. Then the interactions in the system are mediated not by photons, as in the case of the electromagnetic field, but by different type of massless excitations. In the case that the critical point has a quantum origin, and instead of temperature
	certain quantum parameters govern the fluctuations in the system, one speaks of a \textit{quantum critical Casimir effect} \cite{CDT2000,BDT2000}. In addition, systems like liquid $^4$He and liquid crystals, i.e., so-called correlated fluids, exhibit gapless excitations called Goldstone modes \cite{LK91,APP91,LK92,KG99}. These fluctuations lead also to long-ranged forces between the boundaries of the systems,  although such systems are thermodynamically positioned below their respective critical points.  For these cases one speaks of the noncritical or, more generally, the \textit{thermodynamic Casimir effect}. We shall use the latter notion as a general one that encompasses all cases in which the Casimir effect is due to the fluctuations of a certain order parameter. 
	
	Looking on the problem from historical perspective, thirty years after Casimir, in 1978 Fisher and De Gennes \cite{FG78} have shown that a very similar effect exists in fluids. 
	
	As a main set-up for discussing of the thermodynamic Casimir effect, we envisage two material bodies $(1)$ and $(2)$ immersed in a fluid --- see Fig. \ref{Fig:TCE}. They exert an effective force $F^{\rm (1,2)}$ onto each other which is mediated by the fluid. This includes, inter alia, the direct interaction between the material bodies $(1)$ and $(2)$. If the thermodynamic state of the fluid is far away from a bulk phase transition at $T_c$, this force varies slowly and smoothly as function of temperature. Upon approaching $T_c$ of a continuous phase transition, $F^{\rm (1,2)}$  acquires in addition a contribution  $F_\Cas^{\rm(1,2)}$ due to the critical fluctuations of the confined fluid. This additive, singular contribution encompasses both the distortion of the local, (eventually) nonzero order parameter, due to the finite distance between $(1)$ and $(2)$, and the fluctuations of the order parameter. The singular contribution  $F_\Cas^{\rm (1,2)}$ follows by subtracting the smooth background contribution (after extrapolating it to the neighborhood of $T_c$) from $F^{\rm (1,2)}$. This corresponds to the standard procedure for obtaining the singular behavior of thermodynamic quantities such as, e.g., the specific heat (see Ref.  \cite{KD92a}). Upon the above construction, in the disordered phase $F^{\rm(1,2)}$ and  $F_\Cas^{\rm (1,2)}$ vanish in the limit of increasing separation between the bodies $(1)$ and $(2)$. 
	
	It will turn out that the effective force $F_\Cas^{\rm (1,2)}$ between $(1)$ and $(2)$ can be \textit{attractive} or \textit{repulsive}. As expected on general grounds and as already partially outlined                                                                   above, the critical Casimir effect depends on the parameters describing the thermodynamic state of the critical medium, like the temperature and an externally applied field (e.g., pressure, excess chemical potential, magnetic field), as well as on the distance $L$ between $(1)$ and $(2)$, i.e., the observed phenomenon is a \textit{finite size effect}: if $L$ increases the effect and therefore the magnitude of the associated force decreases and eventually vanishes.

	 The $L$-dependence of the force can again easily be determined by \textit{dimensional analysis} alone for the general case of a $d$-dimensional system with a film geometry. Taking into account that the surface area $A$ then is $A\propto L^{d-1}$ and that the energy of the fluctuations $E\propto k_B T$, one concludes that for the thermodynamic Casimir effect near the critical point of the system $F_{\rm Cas}^\parallel \propto k_BT/L^d$. For the $(d=3)$-dimensional system one can write the force at the critical point $T=T_c$ in the form \begin{equation}\label{DimensionsFCas}
		F_{A,{\rm Cas}}^{(\tau)}(T=T_c, L)\simeq8.1\times10^{-3}
		\dfrac{\Delta^{(\tau)}(d=3)}{(L/\mu{\rm m})^{3}}\dfrac{T_{c}}{T_{\rm roon}}\dfrac{{\rm N}}{{\rm m^{2}}},
	\end{equation}
	where $T_{\rm room}=20$ $^\circ$C (293.15 K). Here $\Delta^{(\tau)}$ is the so-called Casimir amplitude that depends on the bulk and surface \textit{universality classes} (see below) of the system and the applied \textit{boundary conditions} $\tau$. For most systems and boundary conditions $\Delta^{(\tau)}(d)={\cal O}(1)$. Thus, when $T_{c}\simeq T_{\rm room}$ the thermodynamic Casimir force and the quantum one, can be of the same order of magnitude, i.e.,
	they both can be essential, measurable and obviously significant at or below the micrometer length scale. Let us stress that $\Delta^{(\tau)}(d)$ can be both \textit{positive and negative}, i.e., $F_{A,{\rm Cas}}^{(\tau)}(T,L)$ can be both \textit{attractive and repulsive}. The accepted  terminology terms the negative force as attractive one. 
	
	Any thermodynamic system, which is of finite extent in \textit{at least one} spatial direction, is called a \textit{finite-size system}.  The corresponding  modification of its phase behavior, compared with its bulk one, is described by \textit{finite-size scaling theory} \cite{Ba83,P90,BDT2000}. Because of this profound interconnection between the theory of the thermodynamic Casimir effect and finite-size scaling theory, we shall summarize some basic knowledge concerning  finite-size scaling theory which is relevant for studying the thermodynamic Casimir effect. We start by recalling some basic properties of critical phenomena in bulk systems. In the vicinity of the bulk critical point  $(T_c,h=0)$ ruled by the temperature $T$ and some external field $h$, the bulk correlation length of the order parameter $\xi$ becomes large, and theoretically diverges: $\xi_t^+\equiv\xi(T\to T_c^{+},h=0)\simeq \xi_0^{+} t^{-\nu}$, $t=(T-T_c)/T_c$, and $\xi_h\equiv\xi(T=T_c,h\to 0)\simeq \xi_{0,h} |h/(k_B T_c)|^{-\nu/\Delta}$, where $\nu$ and $\Delta$ are the usual critical exponents and $\xi_0^{+}$ and $\xi_{0,h}$ are the corresponding nonuniversal amplitudes of the correlation length along the $t$ and $h$ axes.  If in a finite system  $\xi$ becomes comparable to  $L$, the thermodynamic functions describing its behavior depend on the ratio $L/\xi$ and take scaling forms given by the finite-size scaling theory. Further information of the phase transitions and related physical and mathematical problems can be found in \cite{Ba83,P90,BDT2000} and the set of articles on the topic cited therein.

	According to the {\it universality hypothesis}, as formulated by Kadanoff \cite{K71}, ``all (continuous) phase transition 
	problems can be divided into a small number of different classes depending upon the dimensionality of the system and the
	symmetries of the ordered state. Within each class, all phase transitions have identical behavior in the critical region, only the names of thermodynamic variables are changed." 
	All such systems are then part of the same {\it universality class}.
	For example,  we have the Ising universality class characterized by the breaking of the $\mathbb{Z}_2$ symmetry of the original effective Hamiltonian for the scalar order parameter, the $XY$ universality class  with a two-component order parameter and a disordered phase with $O(2)$ symmetry, and the Heisenberg universality class characterized by a vectorial order parameter with an $O(3)$ symmetry. Any of these bulk universality classes is accompanied with a set of surface universality classes, which depend on what  is the behavior of the order parameter near and at a surface(s) of the semi-infinite, or finite system. For a film geometry the accumulated till nowadays
	both experimental and theoretical evidences support the statement that the Casimir force is attractive when the boundary conditions on either plate are the same, or similar, and is repulsive when they essentially differ from each other. For the case of a one-component fluid the last means, e.g., that one of the surfaces adsorbs the liquid phase of the fluid, while the other prefers the vapor phase. 
	
	In the remainder of the current review we will concentrate on the thermodynamic Casimir effect in a system with  $\infty^{d-1} \times L$ film geometry. We envisage a system exposed at a temperature $T$ and to an external ordering field $h$ that couples to its order parameter - density, concentration difference, magnetization $M$, etc. We imagine as  examples simple fluid system at its liquid - vapor critical point, a magnet at the phase transition from paramagnetic to  ferromagnetic state, or a binary liquid mixture with phases $A$ and $B$ near its consolute temperature point, or a binary alloy. Let $(T=T_c, h=0)$ is this bulk critical point in the $(T,h)$ plane. We will consider only the case of an one-dimensional order parameter $\phi\in \mathbb{R}$. The thermodynamic Casimir force $F_{\rm Cas}(T,h,L)$ in such a system is the \textit{excess pressure}, over the bulk one, acting on the boundaries of the finite system, which is due to the finite size of that system, i.e., 
	\begin{equation} \label{Casimir}
		F_{\rm Cas}(T,h,L)= P_L(T,h)-P_b(T,h).
	\end{equation} 
	Here $P_L$ is the pressure in the finite system, while $P_b$ is that one in the infinite system. Let us note that the above definition is actually equivalent to another one which is also commonly used \cite{E90book,K94,BDT2000}
	\begin{equation}
		\label{grandcan}
		F_{\rm Cas}(T,h,L)\equiv-\frac{\partial\omega_{\rm ex}(T,h,L)}{\partial L}=-\frac{\partial\omega_L(T,h,L)}{\partial L}-P_b,
	\end{equation}
	where $\omega_{\rm ex}=\omega_L-L\,\omega_b$ is the excess grand potential per unit area, $\omega_L$ is the grand canonical potential of the finite system, again per unit area, and $\omega_b$ is the density of the grand potential for the infinite system. The equivalence between the definitions, \eq{Casimir} and \eq{grandcan}, comes from the observation that  $\omega_b=- P_b$, and for the finite system with surface area $A$ and thickness $L$ one has $\omega_L=\lim_{A\to \infty} \Omega_L/A$, with $-\partial\omega_L(T,h,L)/\partial L=P_L$. 
	When $F_{\rm Cas}(\tau,h,L)<0$ the excess pressure will be inward
	of the system,, thus it corresponds to an {\it attraction} of the surfaces of the system towards each other, and to a {\it repulsion}, if $F_{\rm Cas}(\tau,h,L)>0$.

	For such a system positioned near its critical point the finite-size scaling theory \cite{C88,BDT2000,Ba83,P90,Ped90,K94,KD92a} predicts:
	
	$\bullet$ For the Casimir force
	\begin{equation}\label{cas}
		F_{\rm Cas}(t,h,L)=L^{-d}X_{\rm Cas}(x_t,x_h);
	\end{equation}
	
	$\bullet$ For the order parameter profile
	\begin{equation}\label{mfss}
		\phi(z,T,h,L)= a_h L^{-\beta/\nu}X_\phi
		\left(z/L,x_t,x_h\right),
	\end{equation}
	where
	$x_t=a_t t L^{1/\nu}$, $x_h=a_h h L^{\Delta/\nu}$.
	In Eqs. (\ref{cas}) and (\ref{mfss}), $\beta $ is the critical exponent for the order parameter, $d$ is the dimension of the system, $a_t$ and $a_h$ are nonuniversal metric factors that can be fixed, for a given system, by taking them to be, e.g., $a_t=1/\left[\xi_0^+\right]^{1/\nu}$, and $a_h=1/\left[\xi_{0,h}\right]^{\Delta/\nu}$, $\beta, \Delta$ and $\nu$ are the so-called critical exponents describing the singular behavior of the physical quantities near the bulk critical temperature $T_c$, and $t=(T-T_c)/T_c$ measures the dimensionless temperature difference away from $T_c$ at which the system is thermodynamically positioned.

	\subsection{Other examples of fluctuation induced forces}
	
	Here we just briefly mention few other fluctuation induced forces for which, due to the space limitations, we will avoid going into any details. 
	
\textit{\textbf{(i) }} Several type fluctuation-induced forces are related to charge fluctuations --- see Refs. \cite{PS1998,AAEH2022,KS52,P89,HL97,HP2005,NDSHP2010,DBWPW2016}.

\textit{\textbf{(ii)}} There are fluctuation induced forces between objects on a fluctuating membrane or on fluid interfaces --- see Refs. \cite{GBP93,BDF2010,LOD2006,OD2008,BRF2011,MVS2012,NWZ2013}. 
	
\textit{\textbf{(iii)}} Fluctuations of fluid velocities \cite{DW2006} as well as fluctuations in electric fields may both give rise to forces acting on the solute particles in colloidal suspensions.

\textit{\textbf{(iv)}} Due to the phonon-mediated interaction between defects \cite{Ro2019,LR2020} in condensed matter systems one studies the phonon Casimir effect.
	
\textit{\textbf{(v)}} There exists also the so-called non-equilibrium thermodynamic (hydrodynamic) Casimir-like effect where correlations in fluids in nonequilibrium or nonequilibrium steady states are of importance --- see Refs. \cite{KOS2013,KOS2014,KOS2015,KOS2016,KOS2016b,AKK2015,RKK2016,RSKK2017}.

\textit{\textbf{(vi)}} Fluctuation-induced Casimir forces in granular fluids have been reported in Ref. \cite{CBMNS2006}. 
	
\textit{\textbf{(vii)}} In nematic liquid crystals the fluctuations of the
	nematic director are responsible for the
	long-ranged nature of the corresponding Casimir force \cite{APP91,ADHPP92,LKS93,ZZ96,ZPZ99,HSD2004,HSD2005,HD2006}.  
	
\textit{\textbf{(viii)}} We mention here also studies of the Casimir effect in active matter  systems \cite{RRR2014,RSKK2017,Kjeldbjerg2021,Tayar2022,BACV2022,Fava2022}. 

In the present review we are mainly concerned with the quantum and the thermodynamic Casimir effect. Since these effects depend on the geometry of the system, there is a plethora of phenomena. In order to keep the volume of the review within reasonable limits, we focus on the  film geometry only.

\section{Casimir versus Helmholtz force}
\label{sec:Casimir-versus-Helmholtz}

When the degrees of freedom can enter and leave the region between the interacting objects one speaks about \textit{Casimir force}. Within the statistical mechanics  one describes the effect within the grand canonical ensemble. In recent Letter  \cite{DR2022} we have introduced the terms of a\textit{ Helmholtz} fluctuation induced force. It is a force in which the order parameter value is fixed \cite{rem}. Such phenomena are then described within the canonical ensemble. 

\subsection{Casimir force and grand canonical ensemble}
\label{sec:GCE}

We consider a finite lattice ${\cal L}\in \mathbb{Z}^d$ with each site ${ \mathbf{r}}$ on the $d$-dimensional lattice embedded with a spin variable $\mathbf{s}_{{ \mathbf{r}}}\in \mathbb{R}^n$. The spins interact with via exchange interactions $J(\mathbf{r},\mathbf{r'})$. 
The corresponding Hamiltonian is
\begin{equation}
	\label{eq:basic-Hamiltonian}
	{\cal H}\left(\left\{\mathbf{s}_i\right\}\right)=-  \sum_{\mathbf{r}, \mathbf{r'}\in {\cal L}} J(\mathbf{r},\mathbf{r'})\; \mathbf{s}_{{\mathbf{r}}}\cdot \mathbf{s}_{{\mathbf{r'}}}.
\end{equation}

In the simplest possible case the spins interact with via a nearest neighbor
exchange interaction $J$. Then the sum runs over nearest neighbor pairs $\langle{\mathbf{r}, \mathbf{r'}}\rangle$ on the lattice.  For the simplicity of the notations, in the remainder we will concentrate on only such type of interactions, if it is not explicitly stated otherwise. For $n=1$ on speaks about Ising type models, for $n=2$ --- for $XY$ ones and for $n=3$ --- for the Heisenberg type models. In the simplest Ising type model the spins can only take the values $+1$ and $-1$.

In the statistical mechanics the systems are described within the so-called Gibbs ensembles of dependent random variables. More specifically, in the grand canonical ensemble, i.e., in the presence of an external bulk field $h$, the Hamiltonian of an Ising type model is
\begin{equation}
	\label{eq:Ham_I}
	{\cal H}_{\rm GCE}\left(\left\{s_{\mathbf r}\right\},h\right) = {\cal H}\left(\left\{s_{\mathbf r}\right\}\right)- h \sum_{\mathbf{r}} s_{\mathbf{r}},
\end{equation}
Then, the partition function of the grand canonical ensemble of a system with $N$ particles is
\begin{equation} 
	\label{eq:ZGCE}
	Z_{\rm GCE}(N,\beta,h)=\sum_{\{s_{\mathbf r}\}}\exp{\left[-\beta H_{\rm GCE}(\{s_{\mathbf r}\},h)\right]}, \qquad  \mbox{where} \qquad \beta=1/(k_B T). 
\end{equation}
Obviously, for the total average magnetization $\overline{M}$ one has 
\begin{equation}
	\label{eq:total_magnetization_GCE}
	\overline{M} \equiv \left \langle \sum_{\mathbf{r}} s_{\mathbf{r}} \right \rangle=\frac{\partial}{\partial (\beta h)} \ln \left[	Z_{\rm GCE}(N,\beta,h)\right].
\end{equation}
Within the grand canonical ensemble the corresponding fluctuation induced force is called the thermodynamic Casimir force. Below follows its definition. 

\subsection{The definition of the thermodynamic Casimir force}

Once the partition function, \eq{eq:ZGCE}, is known, one can determine  the total Gibbs free energy ${\cal F}_{ {\rm tot}}$ via
\begin{equation}
	\label{eq:Gibbs-free-energy}
	\beta {\cal F}_{ {\rm tot}}(N,\beta,h)=-\ln \left[	Z_{\rm GCE}(N,\beta,h)\right],
\end{equation}
and the Casimir force \cite{DD2022}
\begin{equation}
	\label{CasDef}
	\beta F_{\rm Cas}^{(\zeta)}(T,h,L)\equiv- \frac{\partial}{\partial L}f_{\rm ex}^{(\zeta)}(T,h,L),
\end{equation}
where
\begin{equation}
	\label{excess_free_energy_definition}
	f_{\rm ex}^{(\zeta)}(T,h,L) \equiv f^{(\zeta)}(T,h,L)-L f_b(T,h)
\end{equation}
is the so-called excess, over the bulk free energy  $L f_b(T,h)$ density contribution, normalized per area and per $k_B T$.
Here one envisages a system in a film geometry $\infty^{d-1}\times L$, $L\equiv L_\perp$, with boundary conditions $\zeta$ imposed along the spatial direction of finite extent $L$, and with total free energy ${\cal F}_{ {\rm tot}}$.
Here   $f^{(\zeta)}(T,h,L)\equiv \lim_{A\to\infty}{\cal F}_{ {\rm tot}}/A$  is the free energy per area $A$ of the system. 

\subsection{Helmholtz force and canonical ensemble}
\label{sec:CE}

Within the canonical Gibbs ensemble, the Hamiltonian of the model is
\begin{equation}
	{\cal H}_{\rm CE}\left(\left\{s_\mathbf{r}\right\}\right) = {\cal H}\left(\left\{s_\mathbf{r}\right\}\right) ,\; \qquad \mbox{with the constraint} \qquad \sum_{\mathbf{r}} s_{\mathbf{r}}= M,
\end{equation} 
i.e., only configurations with a given fixed value of $M$ are allowed. The statistical sum within this ensemble then is
\begin{equation}
	\label{eq:Z-can-ensemble}
	Z_{\rm CE}(N,\beta,M)=\sum_{\{s_\mathbf{r}\},\sum_{i=1}^N s_\mathbf{r}=M}\exp{\left[-\beta H_{\rm CE}(\{s_\mathbf{r}\})\right]}, \qquad \beta=1/(k_B T).
\end{equation}

Within the canonical ensemble the corresponding fluctuation induced force is called the Helmholtz force \cite{DR2022} (see also \cite{rem}). Below follows its definition. 

\subsection{On the definition of the Helmholtz force}
\label{sec:HF}

From $Z_{\rm CE}(N,\beta,M)$, according to the principle of the statistical mechanics, one can determine the total Helmholtz free energy 
\begin{equation}
	\label{eq:Helmholtz-free-energy}
	\beta {\cal A}_{\rm tot}=-\ln Z_{\rm CE}(N,\beta,M),
\end{equation}
which allows for the determination of a fluctuation-induced force $F_{\rm H}(T,M,L)$ in the fixed $M$-ensemble, i.e., in the $T-M$ ensemble, which we will call \textit{Helmholtz force}.  This can be achieved in a manner similar to the definition of the Casimir force for critical systems in the grand-canonical $T$-$h$ ensemble. Along these lines we define 
\begin{equation}
	\label{HelmDef}
	\beta F_{\rm H}^{(\zeta)}(T,M,L)\equiv- \frac{\partial}{\partial L}a_{\rm ex}^{(\zeta)}(T,M,L),
\end{equation}
where
\begin{equation}
	\label{excess_free_energy_definition_M}
	a_{\rm ex}^{(\zeta)}(T,M,L) \equiv L a_H^{(\zeta)}(T,M,L)-L\, a_H(T,m),
\end{equation}
with $m=\left[\lim_{A\to \infty}(M/A)\right]/L$, while  $a_H(T,M,L)\equiv \left[\lim_{A\to\infty}{\cal A}_{ {\rm tot}}/A\right] /L$ is the Helmholtz free energy density of the finite system, and $a_H(T,m)$ is that one of the infinite system.

In \cite{DR2022}, see also Refs. \cite{RSVG2019,GVGD2016,GGD2017}, it is shown that the so-defined \textit{Helmholtz fluctuation induced force } has a behavior very different from that one of the Casimir force. Explicitly, it is demonstrated that for the Ising chain with fixed $M$ under periodic boundary conditions $F_{\rm H}^{\rm (per)}(T,M,L)$ can, depending on the temperature $T$, be attractive or repulsive, while $ F_{\rm Cas}^{\rm (per)}(T,h,L)$ is only attractive. We note that the issue of the ensemble dependence of fluctuation induced forces pertinent to the ensemble has not, to our knowledge, been fully studied up to now. This issue is by no means limited to the Ising chain, or to the canonical ensemble, and can be addressed, in principle, in \textit{any} model of interest and in \textit{any} ensemble. Indeed the Ising chain case can be viewed as a useful addition to approaches to fluctuation-induced forces in the fixed-$M$ ensemble based on Ginzburg-Landau-Wilson Hamiltonians \cite{RSVG2019,GVGD2016,GGD2017} in which one studied the \textit{canonical Casimir force}. Actually, the "canonical Casimir force", considered there is, in fact, a mathematically equivalent version of the Helmholtz force defined here appropriate to the corresponding models considered there. The statistical-mechanical ensembles are normally equivalent in the thermodynamic limit, but not with respect of the behavior of the finite systems. Thus, the corresponding ensemble dependent fluctuation induced forces is expected to have a different behavior, reflecting the different physical situations they do describe. 

\section{Some results for the Helmholtz force}
\label{sec:Helmholzt-force}

\begin{figure}[!htb]
	\begin{minipage}{0.455\columnwidth}
		\includegraphics[width=\columnwidth]{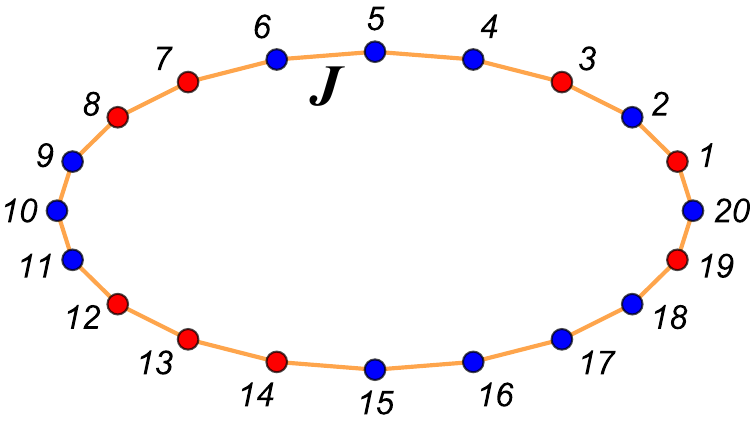}
		\caption{\label{Fig:1d-periodic} 1d-Ising model chain in a ring form. This is equivalent to a system with periodic boundary conditions. In the considered example $M=4$, i.e., the number of "blue" atoms (molecules) is with $4$ more than the "red" ones. One can also think that, say, the blue dots represent spins "up", i.e. $s_i=+1$, while red ones represents spins down, i.e., $s_i=-1$.}
	\end{minipage}\hspace{0.5pc}%
	\begin{minipage}{0.5\columnwidth}
		\includegraphics[width=\columnwidth]{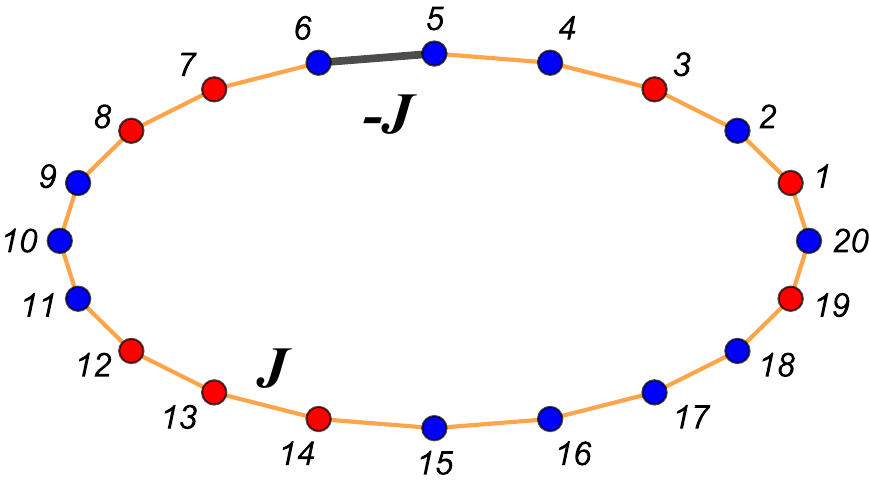}
		\caption{\label{Fig:1d-antiperiodic} 1d-Ising model chain in a ring form and one opposite (defect) bond. This is equivalent to a system with antiperiodic boundary conditions. In the considered example $M=4$, i.e., the number of "blue" atoms (molecules) is with $4$ more than the "red" ones. As in the periodic case, one can think of the blue dots to depict spins "up", i.e. $s_i=+1$, while red ones represent spins down, i.e., $s_i=-1$.}
	\end{minipage} 
\end{figure}

Following Ref. \cite{DR2022}, below we report some exact results for the Helmholtz force for one-dimensional Ising model. Examples of possible configurations with $N=20$ and $M=4$ are shown in Fig. \ref{Fig:1d-periodic} for the case of periodic boundary conditions and Fig. \ref{Fig:1d-antiperiodic} for antiperiodic ones.

The results for both the partition function and the force depend on the boundary conditions.  For the corresponding partition functions one has: 

\textit{i)} for \textit{periodic} boundary conditions
\begin{eqnarray}
	\label{eq:statistical-sum}
	Z^{(\rm per)}(N,K,M) = N e^{K (N-4)} \,_2F_1\left(\frac{1}{2} (-M-N+2),\frac{1}{2} (M-N+2);2;e^{-4 K}\right), \nonumber
\end{eqnarray}
where $_2F_1(\alpha,\beta;\gamma;z)$ is the generalized hypergeometric function \cite{AS}. 
\\
\textit{ii)} for \textit{Dirichlet} (missing neighbors at the both ends of the chain) boundary conditions 
	\begin{eqnarray}
		Z^{(D)}(N,K,M) 
		& = & e^{K (N-1)} \Bigg[2 e^{-2 K} \, _2F_1\left(\frac{1}{2} (-M-N+2),\frac{1}{2}
		(M-N+2);1;e^{-4 K}\right) \\ &&-\frac{1}{2} e^{-4 K} (M-N+2) \, _2F_1\left(\frac{1}{2}
		(-M-N+2),\frac{1}{2} (M-N+4);2;e^{-4 K}\right) \nonumber \\ && + \frac{1}{2} e^{-4 K} (M+N-2) \,
		_2F_1\left(\frac{1}{2} (-M-N+4),\frac{1}{2} (M-N+2);2;e^{-4 K}\right)\Bigg] \nonumber  \label{eq:DDbc-main-text-appendix};
	\end{eqnarray}
	\\
	\textit{iii)} for  \textit{antiperiodic} boundary conditions
	\begin{eqnarray}
		Z^{(\rm anti)}(N,K,M)
		& = &e^{K (N-6)} \left[2 \left(e^{4 K}-1\right) \, _2F_1\left(\frac{1}{2} (-M-N+2),\frac{1}{2}
		(M-N+2);1;e^{-4 K}\right) \right. \nonumber \\ & & \left. +N \, _2F_1\left(\frac{1}{2} (-M-N+2),\frac{1}{2}
		(M-N+2);2;e^{-4 K}\right)\right]  \label{eq:cf33}.
	\end{eqnarray}

If we write $M=mN$ and focus on the case $N \gg 1$, then the exact expressions above approach different forms. In the case of periodic boundary conditions, the partition function becomes
\begin{equation}
	Z^{(\rm per)}_{\rm \lim}(N,K,m) = \frac{2}{N}\frac{e^{NK}x_t}{\sqrt{1-m^2}}I_1(x_t \sqrt{1-m^2}), \label{eq:scalingform}
\end{equation}
where $I_1$ is the modified Bessel function of order 1, and $x_t=Ne^{-2K}$ is the scaling combination $N/\xi_t$, $\xi_t $ being the correlation length \cite{B82} in the vicinity of the zero temperature critical point. This allows us to explore the scaling behavior of thermodynamic quantities close to $T=0$. Limiting forms for the antiperiodic and Dirichlet partition functions can also be obtained.
\begin{figure}[!htb]
	\begin{minipage}{0.455\columnwidth}
		\includegraphics[width=\columnwidth]{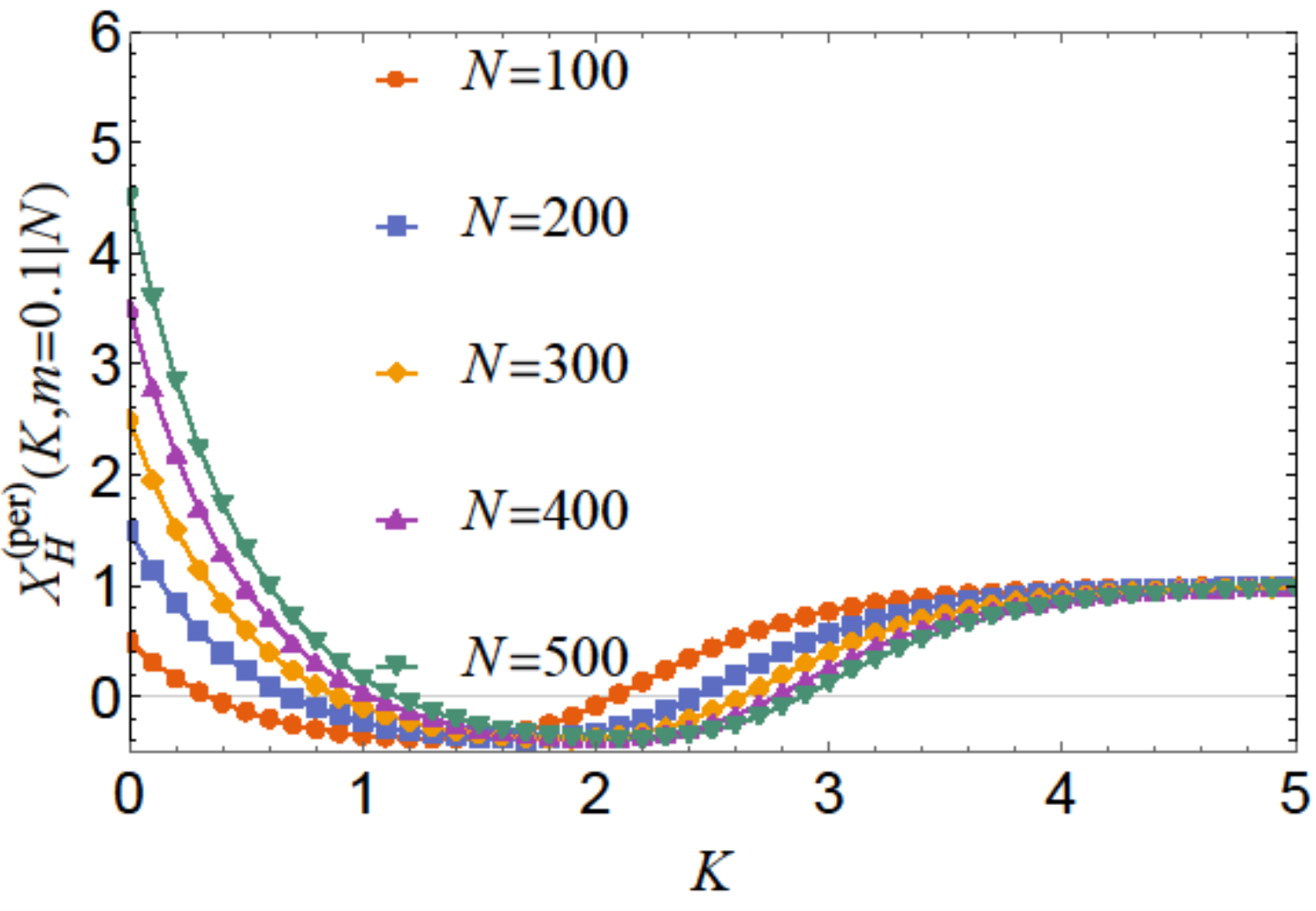}
		\caption{The behavior of the function $X_{\rm H}^{(\rm per)}(K,m|N)$ (see \eq{eq:figeq3}) with $N=100,200, 300, 400$ and $N=500$. We observe that the function is \textit{positive}  for large and  for small enough values of $K$, while being \textit{negative} for relatively moderate  values of $K$, \textit{irrespective} of the value of $N$. The larger $N$, the stronger the repulsion  is for a small enough $K$; the force in the latter regime is strongly repulsive, irrespective on the value of $N$.   }
		\label{fig:Helmholtz2}
	\end{minipage}\hspace{0.5pc}%
	\begin{minipage}{0.5\columnwidth}
		\includegraphics[width=\columnwidth]{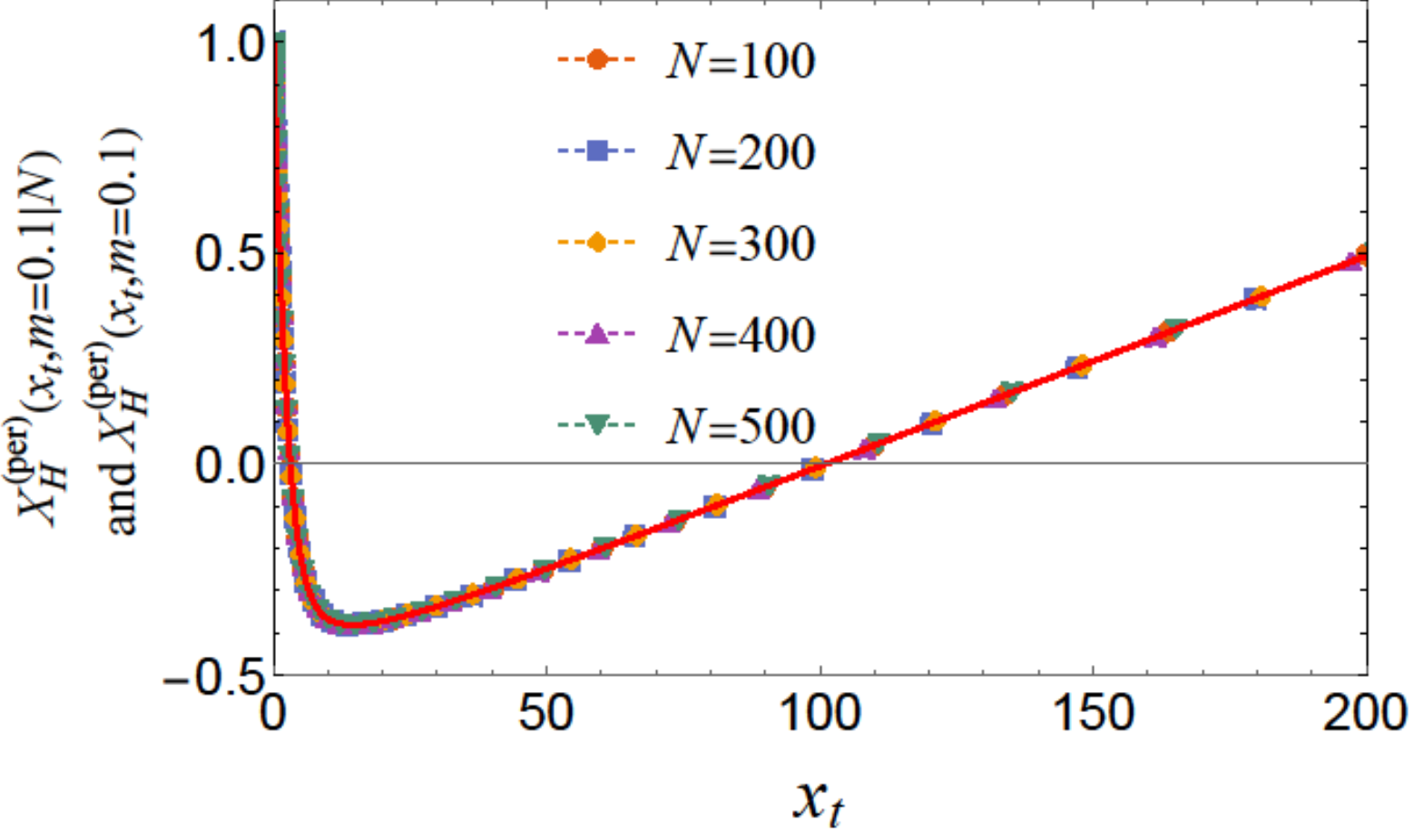}
			\caption{The behavior of the scaling function $X_{\rm H}^{(\rm per)}(x_t,m)$ for $m=0.1$. The inspection of the results obtained numerically from  \eq{eq:statistical-sum} with $N=100,200, 300, 400$ and $N=500$, and that one from \eq{eq:scalingform} demonstrate perfect scaling and agreement between each other. We observe that the function is \textit{positive}  for large values of $x_t$, \textit{negative} for relatively moderate values of $x_t$, and again strongly repulsive for small values of $x_t$. }
		\label{fig:Helmholtz3}
	\end{minipage} 
\end{figure}
From 	\eq{HelmDef} one obtains the fluctuation induced Helmholtz force $F_H^{({\rm per})}(K,m,N)$.
Multiplying the result for $F_H^{({\rm per})}(K,m,N)$ by $N$ provides the function $X_H^{({\rm per})}(K,m|N)$
\begin{equation}
	X_H^{({\rm per})}(K,m|N)=N	F_H^{({\rm per})}(K,m,N) \label{eq:figeq3}.
\end{equation} 
Its behavior is shown in Fig.  \ref{fig:Helmholtz2}. Fig. \ref{fig:Helmholtz2} shows it for $m=0.1$ and $N=100, 200, 300, 400$, and $N=500$. Focusing on the scaling regime ($K$ and $N$ both large compared to 1) we end up with the $N$-independent scaling function $X^{({\rm per})}(x_t,m)$. 
Figure \ref{fig:Helmholtz3} shows the behavior of this quantity as a function of $x_t$ for $m=0.1$.

\subsection{On the connection between the canonical and grand canonical partition functions }
\label{sec:canonical-versus-grand-canonical}

$Z_{\rm CE}(N,\beta,M)$ can be written in the form 
\begin{equation}
	\label{eq:Z-can-ensemble-delta}
	Z_{\rm CE}(N,\beta,M)=\sum_{\{s_\mathbf{r}\}}\exp{\left[-\beta H(\{s_\mathbf{r}\})\right]}\delta\left(\sum_{\mathbf{r}} s_{\mathbf{r}}- M\right).
\end{equation}
Starting from \eq{eq:Z-can-ensemble-delta} and using the identity 
\begin{equation}
	\label{eq:delta-identity}
	\delta(s)=\frac{1}{2\pi}\int_{-\infty}^{\infty} \exp[i\; s\; x] dx
\end{equation}
one consequently obtains
\begin{eqnarray}
	\label{eq:equivalent-representations}
	Z_{\rm CE}(N,\beta,M)&=&\frac{1}{2\pi}\int_{-\infty}^{\infty} dx \sum_{\{s_\mathbf{r}\}}\exp{\left[-\beta H(\{s_\mathbf{r}\})+i x \left(\sum_{\mathbf{r}} s_{\mathbf{r}}- M\right)\right]}\nonumber \\
	& =&\frac{1}{2\pi}\int_{-\infty}^{\infty} dx \exp[-i \; M\; x]  \sum_{\{s_\mathbf{r}\}}\exp{\left[-\beta H(\{s_\mathbf{r}\})+i x \sum_{\mathbf{r}} s_{\mathbf{r}}\right]}\nonumber\\
	& =& \frac{1}{2\pi}\int_{-\infty}^{\infty} dx \exp[-i \; M\, x]\; Z_{\rm GCE}(N,\beta,i\, x).
\end{eqnarray}
Thus
\begin{equation}
	\label{eq:relation-with-grand-canonical}
	Z_{\rm CE}(N,\beta,M)=\frac{1}{2\pi}\int_{-\infty}^{\infty} dx \exp[-i \; M\, x]\; Z_{\rm GCE}(N,\beta,i\, x).
\end{equation}
Using the definitions of the Gibbs free energy density $f(\beta,h,N)$
\begin{equation}
	\label{eq:Gibbs-free-energy}
	Z_{\rm GCE}(N,\beta,h)\equiv \exp[-N \beta f(N,\beta,h)],
\end{equation}
the Helmholtz free energy
\begin{equation}
	\label{eq:Helmholtz-free-energy}
	Z_{\rm CE}(N,\beta,M)\equiv \exp[-N \beta a(N,\beta,h)],
\end{equation}
and introducing the magnetization per particle $m=M/N$, we can rewrite \eq{eq:relation-with-grand-canonical} in the form
\begin{equation}
	\label{eq:relation-Gibbs-Helmholtz-free-enegies}
\exp[-\beta a(N,\beta,m)]=\frac{1}{2\pi}
\int_{-\infty}^{\infty} dx \exp\{-N[\beta f(N,\beta,\;i x)+i \;  m\, x]\}.
\end{equation}
For $N\gg 1$  integrals of this kind can be estimated from the saddle-point
method (see Ref. \cite{F77,Fedoryuk1987}), which in this case reads 
\begin{equation}
	\label{eq:saddle-point-method}
\int_{\mathbb{R}}dx \exp[-N g(x)]\simeq \exp[-N g(x_0)] \sqrt{\frac{2\pi}{N g''(x_0)}}\left(1+{\cal O}(N^{-2})\right); g'(x_0)=0. 
\end{equation}
With the interpretation $i x_0\equiv h$ one obtains
\begin{equation}
	\label{eq:relation-f-a}
	\exp[-N\beta a(N,\beta,m)]\simeq \exp[-N\left(\beta f(N,\beta,h)+m h\right)]\sqrt{\frac{2\pi}{N\beta \chi(N,\beta,h)}}\left(1+{\cal O}(N^{-2})\right),
\end{equation}
where the extreme condition 
\begin{equation}
	\label{eq:extremum}
	\frac{\partial}{\partial x}  [\beta f(N,\beta, i\;x)+i \;m x]=i\;[m+\frac{\partial}{\partial h} [\beta f(N,\beta, h) ]=0
\end{equation}
leads to the standard statistical-mechanical relation $m=-\partial [\beta f(N,\beta, h)]/{\partial h} $, while $\chi(N,\beta,h)\equiv-\partial^2 f/\partial h^2$ is the susceptibility of the finite system in the grand canonical ensemble. Obviously, \eq{eq:relation-f-a} can be rewritten in the form 
\begin{equation}
	\label{eq:f-a-simple}
	\beta a(N,\beta,m)=\beta f(N,\beta,h)+m h-\frac{1}{2N}\ln \frac{2\pi}{N\beta \chi(N,\beta,h)}.
\end{equation}
\eq{eq:f-a-simple} implies that the leading finite-size corrections in the Helmholtz free energy are of the order of $\ln N/N$. These are much stronger correction than for the Gibbs free energy. There it is well known to be exponentially small in $N$ away from the critical temperature \cite{BDT2000}. 

Taking in the r.h.s. of \eq{eq:f-a-simple} the limits $\lim_{N\to \infty}f(N,\beta,h)=f_b(\beta,h)$ and $\lim_{N\to \infty}a(N,\beta,m)=a_b(\beta,m)$, we arrive at the Legendre transformation between the two ensembles, as known from the standard thermodynamics: 
\begin{equation}
	\label{eq:Legendre}
	a_b(\beta,m)=f_b(\beta,h)+h m. 
\end{equation}

Note that the relation given by \eq{eq:relation-with-grand-canonical} can be inverted which leads to
\begin{equation}
	\label{eq:relation-grand-canonical}
	Z_{\rm GCE}(N,\beta,i y)=\int_{-\infty}^{\infty}dM \exp[i \; M\, y]\; Z_{\rm CE}(N,\beta,M),
\end{equation}
or, with $h=iy$, to the self-explained relation 
\begin{equation}
	\label{eq:relation-grand-canonical}
	Z_{\rm GCE}(N,\beta,h)=\int_{-\infty}^{\infty}dM \exp[h\, M]\; Z_{\rm CE}(N,\beta,M)=\int_{-N}^{N}dM \exp[h\, M]\; Z_{\rm CE}(N,\beta,M),
\end{equation}
where we have taken into account that $|M|\le N$. 
As \eq{eq:relation-grand-canonical} implies, the partition functions $Z_{\rm GCE}(N,\beta,h)$ and $Z_{\rm CE}(N,\beta,M)$ are mutually related via an integral transformation. Their finite-size behavior is, however different. Because of that it is reasonable to expect, as it turns out to be the case of the Ising model, that the fluctuation induced forces in a given ensemble are strongly ensemble dependent.

\section{Concluding comments and discussion}
\label{sec:conclusion}

The interest in the fluctuation-induced phenomena in the last years blossomed due to their importance in the rapidly developing field of nanotechnology where, below a micrometer distances, the van der Waals force (vdWF) and QED CF play a dominant role between neutral nonmagnetic objects. The last implies that these forces play a key role in micro- and nano-electromechanical systems (MEMS/NEMS) \cite{CAKBC2001a,DBKRCN2005,BEBDPS2012} operating at such distances. In vacuum, or gas medium, they lead to  irreversible, usually undesirable phenomena, such as stiction (i.e., irreversible adhesion) or pull-in due to mechanical instabilities \cite{BR2001,BR2001a,CAKBC2001}. Indeed, being negligible at macroscopic distances, the Casimir force can become impressively strong at micro- and nano-scales. According to \eq{qCasimirf}, if two perfectly conducting parallel
metal plates are facing each other at a distance of the order of $10$ nm in vacuum and at zero temperature,  the attractive Casimir  force per area, i.e., the Casimir pressure, can be as large as one atmosphere! Obviously, this affects the design and the functioning of devices at these scales. This can be considered as \textit{first fundamental problem of nanotechnology}.  Obviously, such a
large force strongly influences the performance of micro- and nano-machines by
causing stiction, i.e., their moving parts stick together (de facto irreversibly) and stop working.

Closely related to the above  is another troubling effect: when a particle's characteristic size is scaled down below a micrometer the role of its weight becomes negligible. As a result, when one tries to release such a neutral particle from, say, the surface of whatever handling device in air or vacuum, the particle will not drop down under
the gravity but, instead, will stick to the surface due to the effect of the omnipresent vdWF. If, in an attempt to release the particle, one charges the particle, forces vibration of the surface in question, etc., the released particle might move in an uncontrollable way leaving the observation field of the apparatus controlling the performance of the operation. That is the main reason why the handling, feeding, trapping and fixing of micro- and/or nano- particles is still the main bottleneck in micro manufacturing and is far from being solved in a satisfactory fashion \cite{CVP2005}. This can be considered as \textit{second fundamental problem of nanotechnology}.

Thus, formalizing the above, one of the main problems in the micro- and nano-assembly is the precise and reliable manipulation of a micro- or nano-particles that includes moving it from a given starting point, where it is to be taken from, to some end point, when it is to be placed on. In that respect it seems ideal, if one can modify the net force between the manipulated particle and the operating device, sometimes called gripper, in such a way that it is repulsive at short distances between the handling surfaces and the particle and attractive at larger ones. It is clear that the ability to modify the Casimir interaction can strongly influence the development of MEMS/NEMS. Several theorems, however, seriously limit the possible search of repulsive QED CF \cite{KK2006,S2010,RKE2010}. Currently, apart from some suggestions for achieving QED Casimir repulsion in systems out of equilibrium the only experimentally well verified way to obtain such repulsive force is to have interaction between two different materials characterized by dielectric permittivities $\varepsilon^{(1)}$,  and $\varepsilon^{(2)}$, separated by a fluid with permittivity $\varepsilon^{(0)}$, such that \cite{L56,DLP61,LP80} \eq{repulsion_condition} is fulfilled in sufficiently broad frequency range.  In Refs. \cite{MML96,MLB1997,LS2001,LS2002,MCP2009,IIIM2011} QED Casimir repulsion has indeed been observed experimentally for the sphere-plate geometry. In order to minimize the potential negative effects of all possible circuitry at such a small distances and the complications with the isolation, as well as possible problems involving chemical reactions, it seems that one promising strategy for overcoming the obstacles mentioned above is to choose such a fluid as a medium that possesses no free changes dissolved in it and that is inert and do not interact chemically with the materials. That leads us to choose as a fluid, e.g., a nonpolar liquefied noble gas that has critical parameters as close as possible to the normal ones. Such a strategy to overcome the difficulties described above has been suggested in \cite{VD2015,VD2017}.

As far as the Casimir effect is investigated within the framework of statistical mechanics,  the most  results belong to classical systems in the grand canonical ensemble. It is expected that in the future there will be attempts to extend them to dynamical systems, to quantum systems --- including systems with different types of quenches, to systems described by other, say, canonical ensembles, to systems exhibiting disorder, and topological phase transitions --- as well as to a combination of them.  On the example of the Ising model we have outlined the grand canonical ensemble in section \ref{sec:GCE}. We have also pointed out that one can consider ensemble dependent fluctuation induced forces. For example, we have outlined the canonical ensemble in section \ref{sec:CE} and the definition of the corresponding fluctuation induced force in section \ref{sec:HF}. Also there, following Ref. \cite{DR2022}, we have shown that this forces have behaviors quite different from the ones of the Casimir force under the same boundary conditions and under the same geometry - see Figs. \ref{fig:Helmholtz2} and \ref{fig:Helmholtz3}. Let us note, that all the issues studied for the Casimir forces will be objects of investigation in, say, canonical, or micro canonical ensemble.

Let us finish this short review, by noting that the fluctuation induced forces, and especially the Casimir effect and related force is not only a topic of interest for academic investigations. Similar to the QED Casimir effect, for which the first practical applications are under discussion  (see, e.g., Refs. \cite{IMCB2014,FAKA2014,FMRA2014,WDTRRP2016,YHZM2018,PSS2020,MCKBS2021,GCMSM2021,XuGBJL2022,Schmidt2022} and the references  cited therein), also certain applications of the critical Casimir effect have been already considered  (see, e.g., Refs.  \cite{IMC2005,TZGVHBD2011,DLMP2016,NNVKBS2017,GSL2018,MBCHLKS2019,MCSGDV2019,VMKSK2021,SRJRGSBS2021,XiLSL2021,VGCL2021}). 

\ack

The financial support via Grant No
BG05M2OP001-1.002-0011-C02 financed by OP SESG 2014-2020 and by the EU through the ESIFs is gratefully acknowledged.

\section*{References}

\begin{bibliography}{200}
	
	\providecommand{\newblock}{}

\end{bibliography}	

\end{document}